\documentclass[twocolumn,nofootinbib,amsmath,prl,aps,superscriptaddress,tightenlines,preprintnumbers]{revtex4}

\usepackage{graphicx}
\usepackage{braket}
\usepackage[usenames]{color}
\usepackage{latexsym}

\usepackage{mathrsfs}

\usepackage{tikz}
\usetikzlibrary{arrows,decorations.pathmorphing,backgrounds,positioning,fit,petri,automata,shadows,calendar,mindmap,
decorations.markings,calc}

\def\bea{\begin{eqnarray}}
\def\eea{\end{eqnarray}}

\begin{document}

\newcount\hour \newcount\minute
\hour=\time \divide \hour by 60
\minute=\time
\count99=\hour \multiply \count99 by -60 \advance \minute by \count99
\newcommand{\mydate}{\ \today \ - \number\hour :00}

\title{Higgs Decay to Two Photons at One Loop in the Standard Model Effective Field Theory.}

\author{Christine Hartmann and Michael Trott,\\
Niels Bohr International Academy,
University of Copenhagen, \\
Blegdamsvej 17, DK-2100 Copenhagen, Denmark
}

\begin{abstract}
We present the calculation of the $CP$ conserving contributions to $\Gamma(h \rightarrow \gamma \gamma)$, from dimension six operators at one-loop order, in the linear standard model effective field theory.
We discuss the impact of these corrections on interpreting current and future experimental bounds on this decay.
\end{abstract}

\maketitle
\newpage

\paragraph{\bf I. Introduction.}
With the first beams circulating in the Large Hadron Collider (LHC) for Run II, and the anticipated data set that will be obtained in the near future, the utility of precise calculations of the properties of the standard model (SM) Higgs boson is clear. Improving the calculations of Higgs properties in the linear SM effective field theory (SMEFT), allows this generalization of the SM to consistently accommodate more precise constraints, or small deviations in the properties of the Higgs.

The experimental precision with which $\Gamma(h \rightarrow \gamma \gamma)$ is expected to be measured in Run II is projected to be  $ \lesssim 10 \, \%$ with $\int \mathcal{L} \, dt = 300 \,{\rm fb^{-1}}$ of the data \cite{Flechl:2015foa}.
This motivates us to calculate contributions to the process $\Gamma(h \rightarrow \gamma \, \gamma)$ due to dimension six operators at one loop. Such contributions can modify this decay at the few percent level.
We loop improve the $CP$ conserving operators in this paper, as $CP$ odd operators do not interfere with the SM amplitude.

\paragraph{\bf II. Method of calculation.} \label{BFmethod}
We use the background field (BF) method with $R_\xi$ gauge fixing~\cite{'tHooft:1975vy,DeWitt:1967ub,Abbott:1981ke}, and we define counterterm subtractions consistent with the modified minimal subtraction renormalization scheme.
The renormalization is carried out in $d = 4 - 2\epsilon$ dimensions.
The gauge fixing is implemented as in Refs.~\cite{Hartmann:2015oia,Einhorn:1988tc,Denner:1994xt}. We have explicitly checked to see that the dependence on the introduced gauge parameter cancels in the results. The Goldstone bosons of the SM Higgs doublet field, $\phi^\pm, \phi_0$, are defined through the convention
\bea \label{Hexpansion}
 H = \frac{1}{\sqrt{2}}\left( \begin{array}{c}
\sqrt{2} i \phi^+ \\
 h + \bar{v}_T + \delta v + i\phi_0 \end{array}  \right).
 \eea
 Here, $\bar{v}_T$ is the tree level vacuum expectation value (VEV) in the SMEFT, while $\delta v$ is the one-loop contribution to the VEV as defined in Ref.~\cite{Hartmann:2015oia}.
The BF Ward identities are unbroken, resulting in the following relations among the SM counterterms~\cite{Denner:1994xt}:
\begin{align} \label{BFgoodness}
Z_A Z_e &= 1, &
Z_h &= Z_{\phi_{\pm}} = Z_{\phi_0},
\nonumber \\
 Z_W Z_{g_2} &= 1.
\end{align}
This method leads to several technical simplifications \cite{Hartmann:2015oia}. The Higgs boson is treated as a classical external field consistent with a narrow width approximation. Finite terms enter the calculation due to one-loop
renormalization conditions fixing the asymptotic photon and Higgs states, and the gauge couplings. Our results are the one-loop expression for $\Gamma(h \rightarrow \gamma \gamma)$
in the SMEFT due to $CP$ conserving operators, including these finite terms.
\paragraph{\bf III. Dimension six operators in the decay.}
We follow the operator notation and basis of Ref.~\cite{Grzadkowski:2010es} with $\varphi$ exchanged for $H$.
The operators are normalized
with dimensionful Wilson coefficients including a factor of $1/\Lambda^2$, and each gauge field strength multiplies its corresponding gauge coupling.
$\sigma^a$ are the weak isospin Pauli matrices and the $g_{1,2,3}$ are the SM gauge couplings. Here $q,l$ are the left handed $\rm SU(2)_L$ fields, with indices $a,b,c$.
The $\rm SU(2)_L$ generators in the covariant derivative are normalized as $\tau^a = \sigma^a/2$.
Fermion fields are also labelled with the flavour index $p,r,s,t$.
$\tilde{H}_a = \epsilon_{ab} \, H^{\dagger,b}$, with the $\rm SU(2)_L$ invariant tensor defined by $\epsilon_{12} = 1, \epsilon_{ab} = -\epsilon_{ba}$.
We canonically normalize the theory as in Ref.~\cite{Alonso:2013hga} but the corresponding $\bar{g}$ notation is suppressed in this paper.
$N_c$ is the number of colours and $Y_p = m_p \, \sqrt{2}/v$ is the SM Yukawa coupling. We use the normalization $\sigma_{\mu \nu} = i (\gamma_\mu \gamma_\nu - \gamma_\nu \, \gamma_\mu)/2$ for the anti-symmetric tensor.
The bare operators are defined in terms of bare parameters with the $(0)$ labels suppressed on the fields and couplings.
In Ref.~\cite{Hartmann:2015oia} the one-loop contribution due to $\mathcal{O}_{HB}^{(0)},\mathcal{O}_{HW}^{(0)}, \mathcal{O}_{HWB}^{(0)}$
\bea\label{operators1}
\mathcal{O}_{HB}^{(0)} &=& g_1^2 \, H^\dagger \, H \,  B_{\mu \, \nu} \, B^{\mu \, \nu}, \nonumber \\
\mathcal{O}_{HW}^{(0)} &=& g_2^2 \, H^\dagger \, H \,  W^a_{\mu \, \nu} \, W_a^{\mu \, \nu},  \nonumber \\
\mathcal{O}_{HWB}^{(0)} &=& g_1 \, g_2 \, H^\dagger \, \sigma^a H \,  B_{\mu \, \nu} \, W_a^{\mu \, \nu},
\eea
was determined, including the scale independent electroweak finite terms. Here, we complete this calculation to include the full set of operators,
using the renormalization group results of Refs.~\cite{Grojean:2013kd,Jenkins:2013zja,Jenkins:2013wua,Alonso:2013hga}.
Fig.~\ref{direct_diagrams} shows the diagrams that correspond to the remaining ``direct" contributions (denoted as a box in the figure).
Direct contributions refer to effects in the SMEFT that are not only due to a redefinition of parameters and fields.
The canonically normalized effective Lagrangian with mass eigenstate fields is found as in Ref.~\cite{Alonso:2013hga}.

\begin{align}
\nonumber
\end{align}
\begin{widetext}
\begin{align}\label{operators2}
\mathcal{O}^{(0)}_{\substack{eW\\ rs}}&=   g_2 \, \bar{l}_{r,a} \sigma^{\mu\nu} \, e_s \, \tau_{ab}^I \,  H_{b} \, W_{\mu \, \nu}^I, &
\mathcal{O}^{(0)}_{\substack{eB\\ rs}}&=  g_1 \,  \bar{l}_{r,a} \sigma^{\mu\nu} \, e_s  \,  H_{a} \, B_{\mu \, \nu}, &
\mathcal{O}^{(0)}_{\substack{uW\\ rs}} &= g_2 \, \bar{q}_{r,a} \sigma^{\mu\nu} \, u_s \, \tau_{ab}^I \,  \tilde{H}_{b} \, W_{\mu \, \nu}^I, \nonumber \\
\mathcal{O}^{(0)}_{\substack{uB\\ rs}}&= g_1 \, \bar{q}_{r,a} \sigma^{\mu\nu} \, u_s  \,  \tilde{H}_{a} \, B_{\mu \, \nu}, &
\mathcal{O}^{(0)}_{\substack{dW\\ rs}}&=  g_2 \,  \bar{q}_{r,a} \sigma^{\mu\nu} \, d_s \, \tau_{ab}^I \,  H_{b} \, W_{\mu \, \nu}^I, &
\mathcal{O}^{(0)}_{\substack{dB\\ rs}}  &= g_1 \, \bar{q}_{r,a} \sigma^{\mu\nu} \, d_s \,  H_{a} \, B_{\mu \, \nu},  \nonumber \\
\mathcal{O}^{(0)}_{\substack{eH\\ pr}} &=  H^{\dagger} H (\bar{l}_p e_r H) , &
\mathcal{O}^{(0)}_{\substack{uH\\ pr}}  &= H^{\dagger} H (\bar{q}_p u_r \tilde{H}), &
\mathcal{O}^{(0)}_{\substack{dH\\ pr}}   &= H^{\dagger} H (\bar{q}_p d_r H), \nonumber \\
\mathcal{O}^{(0)}_{H}&= (H^{\dagger} H)^3, &
\mathcal{O}^{(0)}_{H \Box } &=  H^{\dagger} H \Box (H^{\dagger} H ), &
\mathcal{O}^{(0)}_{HD}  &= (H^{\dagger}D_{\mu} H)^*(H^{\dagger} D^{\mu} H ), \nonumber \\
\mathcal{O}^{(0)}_W &= g_2^3 \, \epsilon_{abc}W_{\mu}^{a\nu}W_{\nu}^{b\rho}W_{\rho}^{c\mu}.
\end{align}
\end{widetext}
\paragraph{\bf IV. Renormalization.}
The divergences present in the calculation cancel once the SM counterterms and
the operator counterterms are subtracted.
At one loop, the
mixing effects induced are taken into account by
the renormalization matrix $\mathcal{O}_i^{(0)} = Z_{i,j} \, \mathcal{O}_j^{(r)}$,
introduced to cancel the divergences
produced by the new operators.
We use the indices $i,j,k$ for labeling operators on the
dimension six operator basis of the SMEFT.
The one-loop operator counterterm matrix $\mathcal{Z}_{i,j}$ is normalized
as $ Z_{i,j} = \delta_{i,j}  + \mathcal{Z}_{i,j}/16 \, \pi^2 \, \epsilon$. The $3 \times 3$ submatrix for the operators $\mathcal{O}_{HB}, $ $\mathcal{O}_{HW}, $ $\mathcal{O}_{HWB}$ was determined in Ref.~\cite{Grojean:2013kd}
 and incorporated into the full one-loop calculation in Ref.~\cite{Hartmann:2015oia}. On the basis where $ \mathcal{O}_i =  (\mathcal{O}_{HB}, $ $\mathcal{O}_{HW}, $ $\mathcal{O}_{HWB}, $ $\mathcal{O}_W, $ $  \mathcal{O}_{eB}, $ $\mathcal{O}_{eB}^*, $ $\mathcal{O}_{uB}, $ $\mathcal{O}_{uB}^*, $ $\mathcal{O}_{dB}, $
$\mathcal{O}_{dB}^*, $ $\mathcal{O}_{eW}, $ $\mathcal{O}_{eW}^*, $ $\mathcal{O}_{uW}, $ $\mathcal{O}_{uW}^*, $ $\mathcal{O}_{dW}, $ $\mathcal{O}_{dW}^*)$ and $\mathcal{O}_j = ( \mathcal{O}_{HB}, $ $\mathcal{O}_{HW}, $ $\mathcal{O}_{HWB})$,
the remaining counterterm matrix $\mathcal{Z}_{i,j}$ (with $i > 3$) that is relevant is given in Eq.~\ref{countertermmatrix}.
We define $ \mathcal{N}_j = C_i \, \mathcal{Z}_{i,j}/16\pi^2 \epsilon$. The contributions to the $\mathcal{N}_j$ from Eqn.~\ref{countertermmatrix} are denoted $\Delta \mathcal{N}_j$. These terms enter through the effective Lagrangian as a tree level counterterm
%
\bea
\mathcal{L}_{eff}^{tree} & =  h \,\bar{v}_T \,  \left(\Delta \, \mathcal{N}_{HB}  + \Delta \, \mathcal{N}_{HW} - \Delta \, \mathcal{N}_{HWB} \right) e^2 \, A_{\mu\nu}A^{\mu\nu}.
\eea
\begin{widetext}

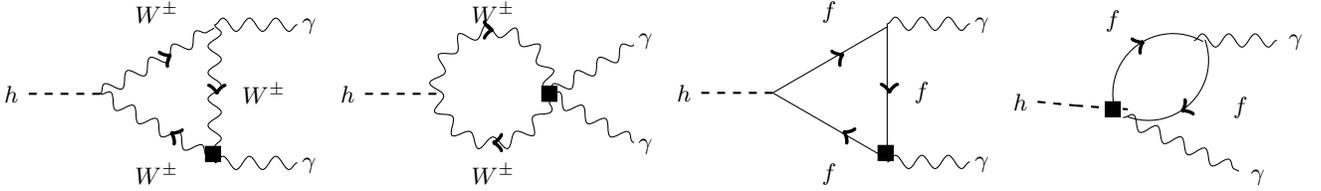
\begin{figure}[!h]
\begin{tikzpicture}[
decoration={
	markings,
	mark=at position 0.55 with {\arrow[scale=1.5]{stealth'}};
}]

\draw  [dashed] [thick] (-1.7,0) -- (-0.75,0);

\draw[decorate,decoration=snake] (180:0.75) -- +(30:1.75) ;
\draw  [->][ultra thick]  (70:0.55)  -- + (30:0.01) ;

 \draw[decorate,decoration=snake] (180:0.75) -- +(330:1.75) ;
\draw  [->][ultra thick]  (290:0.55)  -- + (330:-0.01) ;

\draw[decorate,decoration=snake] (50:1.2) -- +(270:1.75) ;
\draw  [->][ultra thick]  (0:0.8)  -- + (270:0.01) ;

\draw[decorate,decoration=snake] (50:1.2) -- +(0:1.1) ;
\draw[decorate,decoration=snake] (310:1.2) -- +(0:1.1) ;

\filldraw (0.65,-0.9) rectangle (0.85,-0.7);


\node [left][ultra thick] at (-1.7,0) {$h$};
\node [above][ultra thick] at (0,0.8) {$W^{\pm}$};
\node [below][ultra thick] at (0,-0.8) {$W^{\pm}$};
\node [right][ultra thick] at (1,0) {$W^{\pm}$};
\node [right][ultra thick] at (1.8,0.9) {$\gamma$};
\node [right][ultra thick] at (1.8,-0.95) {$\gamma$};

\end{tikzpicture}
\begin{tikzpicture}

\draw  [decorate,decoration=snake] (0,0) circle (0.75);

\filldraw (0.65,-0.1) rectangle (0.85,0.1);


\draw[decorate,decoration=snake] (0:0.75) -- +(30:1.30) ;
\draw[decorate,decoration=snake] (0:0.75) -- +(330:1.30) ;
\draw  [->][ultra thick]  (90:0.85)  -- + (0:0.01) ;
\draw  [->][ultra thick]  (270:0.65)  -- + (0:-0.01) ;


\draw  [dashed] [thick] (-1.7,0) -- (-0.75,0);

\node [left][ultra thick] at (-1.7,0) {$h$};
\node [above][ultra thick] at (0,0.8) {$W^{\pm}$};
\node [below][ultra thick] at (0,-0.8) {$W^{\pm}$};
\node [right][ultra thick] at (1.80,0.7) {$\gamma$};
\node [right][ultra thick] at (1.80,-0.7) {$\gamma$};

\end{tikzpicture}
\begin{tikzpicture}[
decoration={
	markings,
	mark=at position 0.55 with {\arrow[scale=1.5]{stealth'}};
}]

\draw  [dashed] [thick] (-1.7,0) -- (-0.75,0);

\draw(180:0.75) -- +(30:1.75) ;
\draw  [->][ultra thick]  (70:0.55)  -- + (30:0.01) ;

 \draw(180:0.75) -- +(330:1.75) ;
\draw  [->][ultra thick]  (290:0.55)  -- + (330:-0.01) ;

\draw (50:1.2) -- +(270:1.75) ;
\draw  [->][ultra thick]  (0:0.8)  -- + (270:0.01) ;

\draw[decorate,decoration=snake] (50:1.2) -- +(0:1.1) ;
\draw[decorate,decoration=snake] (310:1.2) -- +(0:1.1) ;

\filldraw (0.65,-0.9) rectangle (0.85,-0.7);


\node [left][ultra thick] at (-1.7,0) {$h$};
\node [above][ultra thick] at (0,0.8) {$f$};
\node [below][ultra thick] at (0,-0.8) {$f$};
\node [right][ultra thick] at (1,0) {$f$};
\node [right][ultra thick] at (1.8,0.9) {$\gamma$};
\node [right][ultra thick] at (1.8,-0.95) {$\gamma$};

\end{tikzpicture}
\begin{tikzpicture}

\draw  [dashed] [thick] (-1.7,0) -- (-1.2,-0.05);
\draw  [thick][dashed] (-1.2,-0.05) -- (-0.5,-0.10);

\draw  (0.5,0.8) arc [radius=0.8, start angle=60, end angle= 195];

\draw(-0.5,-0.2) arc [radius=0.8, start angle=250, end angle= 380];

\draw[decorate,decoration=snake] (65:0.9) -- +(0:1.1) ;
\draw[decorate,decoration=snake] (200:0.6) -- +(335:1.7) ;

\draw  [->][ultra thick]  (110:0.85)  -- + (30:0.01) ;
\draw  [->][ultra thick]  (330:0.25)  -- + (210:0.01) ;

\filldraw (-0.6,-0.2) rectangle (-0.8,0);

\node [left][ultra thick] at (-1.7,0) {$h$};
\node [above][ultra thick] at (-0.7,0.8) {$f$};
\node [below][ultra thick] at (1,0.2) {$f$};
\node [right][ultra thick] at (1.5,0.8) {$\gamma$};
\node [right][ultra thick] at (1,-1) {$\gamma$};

\end{tikzpicture}
\caption{Diagrams due to $O_W$,$O_{(e,u,d)B}$,$O_{(e,u,d)W} + H.c$. Mirror graphs are not shown.}
\label{direct_diagrams}
\end{figure}

\end{widetext}
Only a subset of the operators in Eqn.~\ref{operators2} enter through the renormalization group evolution of these operators, which contribute
at tree-level to $\Gamma(h \rightarrow \gamma \gamma)$.
\bea\label{countertermmatrix}
\left( \begin{array}{ccc}
0 & - \frac{15}{2} \, g_2^4  & \frac{3}{2} \, g_2^4   \\
 - (y_l + y_e) \, Y_e  & 0 & - \frac{1}{2} \, Y_e \\
 - (y_l + y_e) \, Y_e^\dagger  & 0 & - \frac{1}{2} \, Y_e^\dagger \\
- N_c  \, (y_q + y_u) \, Y_u & 0  &  \frac{1}{2}  \, N_c \, Y_u \\
- N_c  \, (y_q + y_u) \, Y_u^\dagger & 0  & \frac{1}{2}  \, N_c \, Y_u^\dagger \\
- N_c  \, (y_q + y_d) \, Y_d & 0  & - \frac{1}{2}  \, N_c \, Y_d \\
- N_c  \, (y_q + y_d) \, Y_d^\dagger & 0  & - \frac{1}{2}  \, N_c \, Y_d^\dagger \\
0 & - \frac{1}{2} \, Y_e  & - (y_l + y_e) \, Y_e  \\
0 & - \frac{1}{2} \, Y_e^\dagger  & - (y_l + y_e) \, Y_e^\dagger  \\
 0 & - \frac{1}{2} \, N_c \, Y_u  & N_c\, (y_q + y_u) \, Y_u  \\
  0 & - \frac{1}{2} \, N_c \, Y_u^\dagger  &  N_c\, (y_q + y_u) \, Y_u^\dagger  \\
0 & - \frac{1}{2}\, N_c \, Y_d  & - N_c \, (y_q + y_d) \, Y_d   \\
0 & - \frac{1}{2}\, N_c \, Y_d^\dagger  & - N_c \, (y_q + y_d) \, Y_d^\dagger   \\
 \end{array}
\right),
\eea
The resulting renormalized interactions are
\bea
\mathcal{L}_6^{(0)} &=&  \Big( \mathcal{N}_{HB} \, \mathcal{O}_{HB}^{(r)} +      \mathcal{N}_{HW} \, \mathcal{O}_{HW}^{(r)} +   \mathcal{N}_{HWB}  \, \mathcal{O}_{HWB}^{(r)} \Big) .  \nonumber
\eea
The expression for the counterterm matrix simplifies by noting $(y_l + y_e) - 1/2 = 2 \, Q_\ell,(y_q + y_u) + 1/2 = 2 \, Q_u, (y_q + y_d) - 1/2 = 2 \, Q_d$, and $Q_{\ell,u,d} = \{-1,2/3,-1/3\}$.
Defining $A_{\alpha \, \beta}^{h \, \gamma \, \gamma} =  - 4 \, \left(p_a \cdot p_b \, g^{\alpha \, \beta} - p_a^\beta \, p_b^\alpha \right)$,
we find
\begin{align}\label{treecounter}
\frac{i \mathcal{A}_{\text{tree}}}{A_{\alpha \, \beta}^{h \, \gamma \, \gamma}} & = \frac{-i \, e^2 \, \bar{v}_T}{16 \pi^2 \epsilon} \Bigg( 9  \, g_2^4 \,  C_{W}  + 2 \, Q_\ell \, (C_{\substack{eB \\ rs}} - C_{\substack{eW\\rs}}) \,  [Y_\ell]_{sr}      \Bigg)\nonumber \\
& \hspace{0.2cm}  \frac{-i \, e^2 \, \bar{v}_T \, N_c}{16 \pi^2 \,  \epsilon} \Bigg(2 \, Q_u \, (C_{\substack{uB \\ rs}} + C_{\substack{uW\\rs}}) \,  [Y_u]_{sr} \Bigg) \\
& \hspace{0.2cm} \frac{-i \, e^2 \, \bar{v}_T \, N_c}{16 \pi^2 \, \epsilon} \Bigg(2 \, Q_d \, (C_{\substack{dB \\ rs}} - C_{\substack{dW\\rs}}) \,  [Y_d]_{sr} \Bigg) + H.c. \nonumber
\end{align}
\paragraph{\bf V. Direct contributions.}
In Ref.~\cite{Hartmann:2015oia} the one loop contributions to $\Gamma(h \rightarrow \gamma \gamma)$ due to the operators in Eqn.~\ref{operators1} were determined.
The operators contributing directly in Fig.~\ref{direct_diagrams}
are $\mathcal{O}_W,$ $\mathcal{O}^{(0)}_{(e,u,d)W} + H.c$, $\mathcal{O}^{(0)}_{(e,u,d)B} + H.c$.
Calculating the divergent part of these diagrams, we find exact cancellation with Eq.~\ref{treecounter}.
The finite terms after renormalization are then obtained directly.
\paragraph{\bf VI. Indirect contributions.}
The indirect contributions result from operators redefining fields and parameters within the SM, once the theory is expanded around the VEV. The relevant operators are
$\mathcal{O}^{(0)}_{H},$ $ \mathcal{O}^{(0)}_{H \Box},$ $\mathcal{O}^{(0)}_{HD},$ and $\mathcal{O}^{(0)}_{(e,u,d)H}$.
The divergences associated with these contributions are canceled by the usual SM counterterms, with a reinterpretation of the parameters present in the effective Lagrangian. The
operators $ \mathcal{O}^{(0)}_{H \Box }$ and $\mathcal{O}^{(0)}_{HD}$ contribute to the Higgs and Goldstone boson kinetic terms. Performing the nonlinear field redefinition
\begin{align}\label{fieldredef}
h & \rightarrow h\Big( 1 + (C_{H\Box}  - \frac{1}{4} C_{HD}) \bar{v}_T^2  \Big(1 + \frac{h}{\bar{v}_T} + \frac{h^2}{3 \bar{v}_T^2}  \Big) \Big),
\nonumber \\
\phi_0 & \rightarrow \phi_0 \Big( 1 + \frac{1}{3}(C_{H\Box}  - \frac{1}{4} C_{HD})  \phi_0^2 \Big) ,
\nonumber \\
\phi_+ & \rightarrow \phi_+ \Big( 1 + \frac{1}{2} (C_{H\Box}  - \frac{1}{2} C_{HD})\phi_- \phi_+ \Big) ,
\nonumber \\
\phi_- & \rightarrow \phi_- \Big( 1 + \frac{1}{2} (C_{H\Box}  - \frac{1}{2} C_{HD})\phi_+ \phi_- \Big),
\end{align}
modifies the two derivative interactions in the effective Lagrangian and takes the Higgs kinetic term to canonical form.
The nonlinear field redefinitions on the Goldstone fields
are not required to canonically normalize the theory, but are convenient for later calculations.
This redefinition effects the interactions amongst Higgs and Goldstone bosons.
In the SMEFT, the SM $\lambda$ coupling is shifted as
\begin{align}
\lambda_{\text{SM}} & \rightarrow \lambda \left(1 + \frac{3}{2 \, \lambda} C_H \bar{v}_T^2 - 2C_{H \Box} \bar{v}_T^2 + \frac{1}{2} C_{HD} \bar{v}_T^2  \right) .\nonumber
\end{align}
The vev in the SMEFT is also redefined compared to the SM. We define the vev through the renormalization condition that the one point function of $h$ vanishes to one loop.
The tree level redefinition of the vev is $v_{\text{SM}} \rightarrow \bar{v}_T = v_{\text{SM}} \Big( 1 - \frac{3}{8 \, \lambda} C_H v_{\text{SM}}^2 \Big)$.
The one-loop contribution to the VEV is given by $\delta v$ in Eq.~(\ref{Hexpansion}).
Finally, the effective Yukawa coupling also differs from the SM one. The canonical coupling of the Higgs to the fermion fields in the SMEFT, defined through the convention
$\mathcal{L} = - h \bar{\psi'}_r [\mathcal{Y}]_{rs} P_L \, \psi_s+ H.c.$ where $\psi' = \{u,d,e \}$ and $\psi = \{q,l\}$,  is \cite{Alonso:2013hga}
\bea
[\mathcal{Y}]_{rs} =  \frac{[M_{\psi'}]_{rs}}{\bar{v}_T} \Bigg[1 + \bar{v}^2_T \left(C_{H \Box} -\frac{1}{4} C_{HD} \right)\Bigg] - \, \frac{\bar{v}^2_T}{\sqrt{2}} C^\star_{\substack{\psi' H \\sr}}. \nonumber
\eea
Here, $[M_{\psi'}]_{rs}$ is the mass matrix of the fermions in the SMEFT, as defined in Ref.~\cite{Alonso:2013hga}.
As indirect contributions introduce a finite renormalization (i.e., redefine with a rescaling) of the SM results, these terms are expressible in terms of the well-known functions $A_{1,1/2}$~\cite{Ellis:1975ap,Shifman:1979eb,Bergstrom:1985hp}, where
\bea
A_1(\tau_p) &=&2 + 3 \tau_p \left[1+ (2- \tau_p)f(\tau_p)\right], \\
A_{1/2}(\tau_p)&=&- 2 \tau_p \left[1+ (1- \tau_p)f(\tau_p)\right],
\eea
where $\tau_p = 4 m_p^2/m_h^2$ and
\bea
f (\tau_p) =
\begin{cases}
   \arcsin^2 \sqrt{1/\tau_p} ,&  \tau_p \ge 1\\
    - \frac{1}{4} \left[\ln \frac{1 + \sqrt{1- \tau_p}}{1 - \sqrt{1- \tau_p}} - i \pi \right]^2,              & \tau_p <1.
\end{cases}
\eea
\paragraph{\bf VII. Gauge fixing effects.}
Gauge fixing (GF) has some subtleties in the SMEFT compared to the SM~\cite{Hartmann:2015oia}.
These subtleties are due to the presence of field redefinitions in the SMEFT taking the theory to canonical form.
We apply the nonlinear field redefinitions to the $R_\xi$ GF terms finding the relevant interaction terms,
\begin{align}
\mathcal{L}_{GF} & = - \frac{1}{2} \, g_2^2 \, \xi \, \hat{h} \, \bar{v}_T^3 \, \phi_+ \, \phi_-\left( C_{H\Box} - \frac{1}{4} \, C_{HD} \right) + ...
\end{align}
Here, $\hat{h}$ is the classical background Higgs field in the BF method. We also obtain the
ghost terms from the variation of the GF terms
\bea
\mathcal{L}_{u} = -\frac{1}{2} \, g_2^2  \, \xi \,  \bar{v}_T^3 \, \Big(C_{H\Box} - \frac{1}{4}C_{HD}\Big) \, \hat{h} \, u_{\pm}  \, \bar{u}_{\mp}
\eea
where $u_{\pm}$ are ghost fields. With these redefinitions, the indirect diagrams to calculate are of the same form as in the SM. We have explicitly performed
this calculation with these field redefinitions imposed, finding exact cancellation of the gauge parameter as a cross-check.
Note that the $W^\pm \, \phi^\mp \, \hat{A}$ SM coupling vanishes in the background field method (BFM).
\paragraph{\bf VIII. One-loop $\Gamma(h \rightarrow \gamma \gamma)$ in the SMEFT.}
The result for the impact of $CP$ conserving dimension six operators on the $h \rightarrow \gamma \gamma$ amplitude at one loop in the SMEFT is given as follows.
We define the new physics amplitude as being composed of the individually gauge invariant components $C^{1,NP}_{\gamma \, \gamma}$
and $f_i$ as\footnote{Passarino-Veltman decompositions~\cite{Passarino:1978jh} and the tools Form, FormCalc and FeynCalc~\cite{Vermaseren:2000nd, Hahn:1998yk,Mertig:1990an} have been implemented to carry out the calculations and independent checks of the results.}
\bea
\mathcal{A}^{NP} = \left(C^{1,NP}_{\gamma \, \gamma}   +  \frac{C^{NP}_i \, f_i}{16 \, \pi^2}\right) \, \bar{v}_T \, e^2 \, \mathcal{A}_{\alpha \, \beta}^{h \, \gamma \, \gamma}.
\eea
The index $i$ runs over all of the operators in Eqs.~(\ref{operators1}, \ref{operators2}) other than $O_{HB}$ and $O_H$.
For operators with Hermitian conjugates, the expression that appears multiplying an $f_i$ is ${\rm{Re}}(C_i)$, when considering the $CP$ conserving contributions to $\Gamma(h \rightarrow \gamma \gamma)$.
The coefficient $C^{1,NP}_{\gamma \, \gamma}$ is the one-loop improved set of Wilson coefficients that corresponds to $h \rightarrow \, \gamma \, \gamma$ at tree level.
The tree level expression for this effective Wilson coefficient, in this basis, is given by
\bea
C^{0,NP}_{\gamma \, \gamma} = C_{HW}+ C_{HB} - C_{HWB}.
\eea
The tree level impact of these operators was studied in many works, with Ref.~\cite{Manohar:2006gz} being easily comparable to these results.
The one-loop improvement of this coefficient is scheme dependent. The VEV has to be defined at one loop, as do the external Higgs boson (through $\delta R_h$), the photon states
and the gauge couplings. The renormalization condition for the electromagnetic coupling is fixed in the Thomson limit.
The finite terms at one loop in the definition of the electric charge and the photon wave function residue at the physical pole cancel~\cite{Hartmann:2015oia}
for this process due to the unbroken Ward identities in the BFM; see Eq.~(\ref{BFgoodness}).
This result follows from the theoretical developments in the remarkable Refs.~\cite{Einhorn:1988tc,Denner:1994xt,Denner:1991kt}.
Note that the gauge dependence present in $\delta R_h, \delta v$ and the remaining terms multiplying $C^{0,NP}_{\gamma \, \gamma}$ cancels in the sum contributing
to the amplitude. The one-loop expression is \cite{Hartmann:2015oia}
\bea
C^{1,NP}_{\gamma \, \gamma} &=& C^{0,NP}_{\gamma \, \gamma} \left(1 + \frac{\delta R_{h}}{2} + \frac{\delta \, v}{\bar{v}_T} + (\sqrt{3}  \, \pi  -6) \, \frac{\lambda}{16 \, \pi^2} \right) \nonumber \\
&+& \frac{C^{0,NP}_{\gamma \, \gamma}}{16 \, \pi^2}\left(\frac{g_1^2}{4}+\frac{3 \, g_2^2}{4}+ 6 \, \lambda \right) \log \left( \frac{m_h^2}{\Lambda^2}\right)  \\
&\, & \hspace{-1.5cm} + \frac{C^{0,NP}_{\gamma \, \gamma}}{16 \, \pi^2} \, \Bigg(
 \frac{g_1^2}{4} \, \, \mathcal{I} [m_Z^2] + \left(\frac{g_2^2}{4}+\lambda \right)\, (\mathcal{I} [m_Z^2] + 2 \, \mathcal{I} [m_W^2])\Bigg) . \nonumber
 \eea
Here we have renormalized the theory at the scale $\mu^2 = \Lambda^2$, the cutoff scale of the SMEFT. This means that the
constraints on Wilson coefficients can be directly interpreted as constraints on the underlying theory generating the operators at the scale $\mu \sim \Lambda$.
 The expressions for $\delta R_{h}$ and $\delta \, v$ are somewhat lengthy and are given directly in
Ref.~\cite{Hartmann:2015oia} for $\xi = 1$. The expression for $\mathcal{I}[m_p]$ for $\tau \ge 1$ is
\bea
\mathcal{I}[m_p] & \equiv \log(\frac{\tau_p}{4}) + 2 \sqrt{\tau_p -1} \, \arctan \left(\frac{1}{\sqrt{\tau_p -1}} \right) - 2.
\eea
The remaining $f_i$'s are as follows. We find
\bea
f_{HWB} &=&  \left(- 3 \, g_2^2 +4 \, \lambda  \right) \,  \log \left( \frac{m_h^2}{\Lambda^2}\right) + (4 \lambda - g_2^2) \, \mathcal{I} [m_W^2] \nonumber \\
&-& 4 \, g_2^2 \, \mathcal{I}_y [m_W^2] - 2 \, g_2^2 \left[1+ \log \left( \frac{\tau_{W}}{4}\right) \right] \nonumber \\
&+& e^2 \, (2+ 3 \, \tau_{W})+ 6 \, e^2 (2 - \tau_{W}) \, \, \mathcal{I}_y [m_W^2].
\eea
Here the expression for $ \mathcal{I}_y[m_p] $ for $\tau \ge 1$ is
\bea
\mathcal{I}_y [m_p] \equiv \frac{\tau_p}{2} \arcsin^2 (1/\sqrt{\tau_p}).
\eea
We also find that
\bea
f_{HW} &=& - g_2^2 \Bigg[3 \, \tau_{W} + \left(16 - \frac{16}{\tau_{W}}- 6 \, \tau_{W}\right) \, \mathcal{I}_y [m_W^2] \Bigg], \nonumber \\
f_{W} &=& - 9 \, g_2^4 \,  \log \left( \frac{m_h^2}{\Lambda^2}\right) - 9 \, g_2^4 \,  \mathcal{I} [m_W^2] - 6 \, g_2^4 \,  \mathcal{I}_y [m_W^2] \nonumber \\
&+& 6  \, g_2^4 \,  \mathcal{I}_{xx} [m_W^2] \, \left(1-1/\tau_{W}\right) - 12 \, g_2^4,
\eea
where, for $\tau \ge 1$,
\bea
\mathcal{I}_{xx}[m_p] & \equiv \frac{\tau_p}{\sqrt{\tau_{p} - 1}} \,  \arctan \left(\frac{1}{\sqrt{\tau_{p} - 1}}\right).
\eea
The functions $f_{HW}$ and $f_{W}$ correspond to diagrams with only spin one SM states (in the loops). However, these contributions are not proportional to $A_1$,
which is often termed the ``spin one-loop function" for $h \rightarrow \gamma \, \gamma$. Such terminology is misleading when considering the effective theory generalization of the SM.
The reason the loop functions differ is due to the higher derivative interactions present in the SMEFT, that are forbidden by the (usual $D \leq 4$) renormalizability of the SM.
An interesting consequence of  the different loop functions is that a redefinition of the $W$ mass in the SMEFT as an input parameter is not expected
to absorb the $f_{HW}$ contribution to the amplitude. Note also that the $W$ mass is unchanged by $C_{HW}$ when taking the
theory to canonical form, as in Ref.~\cite{Alonso:2013hga}.
For the dipole leptonic operators, we find,
\bea
f_{\substack{eB \\ ss}}  &=& 2 \, Q_\ell \, [Y_\ell]_{ss}  \left[-1 +  2 \, \log \left(\frac{\Lambda^2}{m_h^2} \right) +  \, \log \left(\frac{4}{\tau_{s}} \right) \right] \nonumber \\
&-& 2 \, Q_\ell \, [Y_\ell]_{ss} \, \Big[ 2 \, \mathcal{I}_{y} [m_s^2] + \mathcal{I}[m_s^2] \Big].
\eea
where the dipole contributions for the quarks follow trivially. Furthermore, we also find that $f_{\substack{eW \\ ss}} =  - f_{\substack{eB \\ ss}}$. Here, $s=\{1,2,3\}$ sums over the flavors of the leptons. The remaining dipole $f_i$'s are obtained by obvious replacements following the structure of the terms
present in Eq.~(\ref{treecounter}).
The indirect contributions are
\bea
[Y_e]_{ss} \, f_{\substack{eH \\ ss}} &=& \frac{Q_\ell^2}{2} A_{1/2}(\tau_s),  \quad
[Y_u]_{ss} \, f_{\substack{uH \\ ss}} = N_c \, \frac{Q_u^2}{2} A_{1/2}(\tau_s),  \nonumber \\
\, [Y_d]_{ss} \, f_{\substack{dH \\ ss}} &=& N_c \, \frac{Q_d^2}{2} A_{1/2}(\tau_s),  \nonumber \\
f_{H \Box} &=& -\frac{Q_\ell^2}{2} A_{1/2}(\tau_p) - N_c \, \frac{Q_u^2}{2} A_{1/2}(\tau_r), \nonumber \\
 &-&  N_c \, \frac{Q_d^2}{2} A_{1/2}(\tau_s) - \frac{1}{2} \, A_1(\tau_{W}),
\eea
and $f_{HD}  = -f_{H \Box} /4$. Here $p,r,s$ run over $1,2,3$ as flavor indices. We have checked to see that we agree with the indirect corrections also reported in
Ref.~\cite{Ghezzi:2015vva}.
Only these indirect results and $C^{0,NP}_{\gamma \, \gamma}$ are included in eHDECAY \cite{Contino:2014aaa}.
\paragraph{\bf VIII. Numerical results.}
The contribution of the one-loop improved amplitude to $\Gamma(h \rightarrow \gamma \gamma)$ is through
\bea
\Gamma(h \rightarrow \gamma \gamma)_{\text{SMEFT}} = \Gamma(h \rightarrow \gamma \gamma)_{\text{SM}} \Bigg|1 - \frac{8 \, \pi^2 \bar{v}_T \, \mathcal{A}^{NP} }{I^\gamma \, e^2 \, \mathcal{A}_{\alpha \, \beta}^{h \, \gamma \, \gamma}} \Bigg|^2
\eea
where $4 \, I^\gamma = A_1(\tau_{W}) + \sum_i \, N_c \, Q_i^2 \, A_{1/2}(\tau_{i})$ is the SM contribution, and the sum over $i$ is over all fermions.
Note that at next-to-leading order, some contributions do not have the form of the SM amplitude and are "nonfactorizable"; for this reason the SM amplitude explicitly divided
out of the second term above will not, in general, cancel.
Experimental constraints on SMEFT parameters is a subject of intense study; however, fully general
global fits of all of the coefficients that are present in $ \mathcal{A}^{NP}$ do not exist.
Furthermore, fits generally are determined neglecting the theoretical error in the SMEFT itself. This is a serious challenge to the consistency of constraints \cite{Berthier:2015oma} when they rise above the percent level (on $C_i \bar{v}_T^2$), for cutoff scales in the $\rm TeV$ range.
As such, we consider the case of unknown $C_i \sim 1$ and vary the unknown parameters over $0.8 \leq \Lambda \leq 3$ in ${\rm TeV}$ units.
Note that $\bar{v}_T^2/(0.8 \, {\rm TeV})^2 \sim 0.1$.

The one-loop improvement of $\mathcal{A}^{NP}$  in the SMEFT can impact the interpretation of experimental constraints on the process $\Gamma(h \rightarrow \gamma \gamma)$
in an interesting manner. Taking $\kappa_\gamma$ from Ref.~\cite{ATLAS-CONF} to be $0.93^{+0.36}_{-0.17}$, and neglecting light fermion ($m_f < m_h$) effects for simplicity, one finds
the one $\sigma$ range
\bea\label{masternumerics}
\frac{-0.02}{4} \, \leq \left(\hat{C}^{1,NP}_{\gamma \, \gamma}   +  \frac{\hat{C}^{NP}_i \, f_i}{16 \, \pi^2}\right) \, \frac{\bar{v}^2_T}{\Lambda^2} \leq \frac{0.02}{4}.
\eea
Here, the hat superscript denotes that the scale $1/\Lambda^2$ has been factored out of a Wilson coefficient.
The difference in the mapping of this constraint to the coefficient of $C^{0,NP}_{\gamma \, \gamma}$ at tree level, and at one loop,
should be corrected as if an inferred coefficient of $C^{0,NP}_{\gamma \, \gamma}$ is to be used in another process as a signal or constraint, or to map to an underlying
model.\footnote{Possibly through an effective bound on a pseudo-observable \cite{Passarino:2012cb,Gonzalez-Alonso:2014eva}}
Interpreting a measured deviation from the SM, in terms of the underlying theory, also requires that this correction is taken into account.
The correction proportional to $C^{0,NP}_{\gamma \, \gamma}$ is dominated by the scheme dependent effects
defining the VEV in $\delta v$, and it is below the few percent level for $\Lambda$ in the range $\left[0.8, \, 3 \right] \, {\rm TeV}$.
\footnote{In these illustrative numerical results,
we are assuming that the residual gauge dependence in $\delta v$ is canceled by a one loop extraction of the vev in muon decay.}

The corrections to the inferred bound on $\hat{C}^{0,NP}_{\gamma \, \gamma}$ due to $f_{i}$ are of interest. To determine this correction we determine the percentage change on the inferred
value of the bounds of $\hat{C}^{0,NP}_{\gamma \, \gamma}$ using Eq.~(\ref{masternumerics}), while shifting the quoted upper and lower bounds by the SMEFT perturbative correction. The
envelope of the two percentage variations on the bounds is quoted in the form $\left[ ,\right]$, for values of $\Lambda$ varying from $\left[0.8, \, 3\right]{ \rm TeV}$.
\footnote{The shifts underestimate
the total impact of NLO corrections to relations between observables.
The effect of one loop corrections in a process using a bound on $C^{0,NP}_{\gamma \, \gamma}$ is neglected.}
For one specific choice of signs for $C_i$ and $\kappa_{\gamma} = \frac{\Gamma(h \rightarrow \gamma \, \gamma)_{\text{SMEFT}}}{\Gamma(h \rightarrow \gamma \, \gamma)_{\text{SM}}}$, we find the results
\bea
[\Delta \hat{C}^{0,NP}_{\gamma \, \gamma}]_{HWB} \sim \left[9, 1 \right]  \% \times 4 \hat{C}_{HWB}.
\eea
The effects of $f_{HW}$ and $f_{W}$ dependence on the inferred bound on $C^{0,NP}_{\gamma \gamma}$ are
\bea
[\Delta \hat{C}^{0,NP}_{\gamma \, \gamma}]_{HW} &\sim& \left[6, 0.1 \right]  \% \times 4 \hat{C}_{HW}, \\
\left[ \Delta \hat{C}^{0,NP}_{\gamma \, \gamma} \right]_{W} &\sim& \left[21, 2 \right]  \% \times 4 \hat{C}_{W}.
\eea
The dependence on the cut off scale in $f_{W}$ is $\sim \log(\Lambda^2)  (1/\Lambda^2)$, while it is $\sim 1/\Lambda^2$ for $f_{HW}$.
The impact of the indirect contributions is
\bea
[\Delta \hat{C}^{0,NP}_{\gamma \, \gamma}]_{tH} &\sim& \left[6, 0.1 \right]  \% \times 4 \,{\rm Re} (\hat{C}_{\substack{uH \\ 33}}), \\
\left[\Delta \hat{C}^{0,NP}_{\gamma \, \gamma} \right]_{H \Box}  &\sim& \left[13, 1 \right]  \% \times 4 \hat{C}_{H \Box}, \\
\left[\Delta \hat{C}^{0,NP}_{\gamma \, \gamma} \right]_{HD} &\sim&  \left[5, 0.1 \right]  \% \times 4 \hat{C}_{HD}.
\eea
Here, we have considered only the top quark coupling effects in $A_{1,1/2}$, for simplicity.
Finally, consider the effect of the top dipole moment operators, which yields
\bea
[\Delta \hat{C}^{0,NP}_{\gamma \, \gamma}]_{dipole} \sim \left[17, 3 \right]  \% \times 4 \left({\rm Re}(\hat{C}_{\substack{uB \\ 33}} ) + {\rm Re}(\hat{C}_{\substack{uW \\ 33}} )\right),
\eea
for $\Lambda$ in the range $\left[0.8, \, 3 \right] \, {\rm TeV}$, falling as $\sim \log(\Lambda^2)  (1/\Lambda^2)$. It
is interesting to note that Ref.~\cite{Kamenik:2011dk} quotes constraints on the dimensionful suppression scale of the dipole operators
$\sim {\rm TeV}$, as considered here. Adding all of the $f_i$ perturbative corrections in quadrature, assuming $\hat{C}_i \sim 1$,
and further correcting the experimental bound for the scheme dependent shift $C^{1,NP}_{\gamma \, \gamma} \rightarrow C^{0,NP}_{\gamma \, \gamma}$, one finds the net
impact of one-loop corrections due to higher dimensional operators on the bound of the tree level Wilson coefficient $
\Delta_{\text{quad}} \,  \hat{C}^{0,NP}_{\gamma \, \gamma} \sim \left[100, 10 \right]  \%.
$ Similarly, CMS reports $\kappa_{\gamma}= 0.98^{+0.17}_{-0.16}$~\cite{Khachatryan:2014jba}, which gives
$
\Delta_{\text{quad}} \,  \hat{C}^{0,NP}_{\gamma \, \gamma} \sim \left[160, 20 \right]  \% .
$%
It is possible that these corrections could add up in a manner that is not in quadrature, as this depends on the
unknown $C_i$ values. In some cases of weakly coupled and renormalizable UV scenarios,
the contributions from  $C_{\gamma \, \gamma}$, $C_W$, $C_{HWB}$, $C_{tB}$, and $C_{tW}$ are all expected to be suppressed by a further loop factor, as the arguments
of Ref.~\cite{Arzt:1994gp} apply. Then the indirect contributions are expected to be the largest effects. No general statement of this form can be made for strongly interacting theories,
or cases where EFTs are present in the UV \cite{Jenkins:2013fya}.

We emphasize that the impact of the one-loop corrections listed above is on {\it current} experimental bounds of $\Gamma(h \rightarrow \gamma \gamma)$.
If these effects are simply neglected, then it is necessary to add a {\it theoretical error for the SMEFT} to the experimental error, which can already be dominant
in precisely measured processes~\cite{Berthier:2015oma}. As the experimental precision of the measurement of $\Gamma(h \rightarrow \gamma \gamma)$ increases, the impact of the neglected corrections
directly scales up. Repeating the exercise above with a chosen $\kappa^{\text{proj:RunII}}_\gamma = 1 \pm 0.045$, consistent with projected future bounds (CMS - scenario II~\cite{Flechl:2015foa,CMS:2013xfa})
$
(\Delta_{\text{quad}} \, \hat{C}^{0,NP}_{\gamma \, \gamma})^{\text{proj:RunII}} \sim 4 \left[167, 21 \right]  \%.
$
High luminosity LHC runs are further quoted to have a sensitivity between $2 \%$ and $5 \%$ in $\kappa_\gamma$  \cite{Dawson:2013bba}. Choosing
 $\kappa^{\text{proj:HILHC}}_\gamma = 1 \pm 0.03$, one finds
$
(\Delta_{\text{quad}} \, \hat{C}^{0,NP}_{\gamma \, \gamma})^{\text{proj:HILHC}} \sim 4 \left[250, 31 \right]  \%.
$
The Snowmass working group report~\cite{Dawson:2013bba} further quotes TLEP projections with a potential future sensitivity  $\kappa^{\text{proj:TLEP}}_\gamma = 1 \pm 0.0145$.
Performing the same exercise with this projection
$
(\Delta_{\text{quad}} \, \hat{C}^{0,NP}_{\gamma \, \gamma})^{\text{proj:TLEP}} \sim 4 \left[513, 64 \right]  \%.
$

{\bf VI. {\bf Conclusions.}}\label{conclusions}
We have presented the one-loop result for $\Gamma(h \rightarrow \gamma \gamma)$ due to $CP$ conserving operators of dimension six in the SMEFT.
The theoretical developments reported in this Letter allow small deviations in the rate of the Higgs decay to two photons to be consistently mapped to
underlying new physics models with states that have TeV mass scales. Such models are of great interest as they might
provide a more compelling -- and less UV sensitive -- description of electroweak symmetry breaking. The results of Run I at the LHC also indicate
that there is (probably) a mass gap between the electroweak scale and an UV origin of electroweak symmetry breaking, indicating that this mapping
must be done if any deviation is discovered in this decay. The calculation reported here can also
be used as a model to loop improve many other processes in the SMEFT.
The loop corrections reported here are already relevant to precisely interpret experimental bounds on
Higgs properties, and should not be neglected in future studies as they are now known.

{\bf Acknowledgements}
M.T. acknowledges generous support by the Villum Fonden and partial support by the Danish National Research Foundation (DNRF91).
The project leading to this application has received funding from the European Union's Horizon 2020 research
and innovation programme under the Marie Sklodowska-Curie grant agreement No 660876, HIGGS-BSM-EFT.
C.H. thanks G. Heinrich for the introduction to the technical tools used throughout the calculations of this paper.
We thank L. Berthier for comments on the manuscript. Note added: We have corrected typos 
in the text pointed out in Ref.~\cite{Dedes:2018seb}, which otherwise confirmed the results of this work
up to an (apparent) numerical scheme dependence. The same results in  Ref.~\cite{Hartmann:2015oia} are
unaffected by the (corrected) typo in the $I^\gamma$ definition,
despite an inaccurate statement in Ref.~\cite{Dedes:2018seb} [arXiv v1] to the contrary.
Ref.~\cite{Hartmann:2015oia} agrees with Ref.~\cite{Manohar:2006gz} in the normalization of its SMEFT corrections.
We also note that Ref.~\cite{Ghezzi:2015vva} was partially compared to
in this work, finding agreement, and Ref.~\cite{Ghezzi:2015vva} also reported a past comparison and agreement with Ref.~\cite{Hartmann:2015oia},
prior to Ref.~\cite{Dedes:2018seb} [arXiv v1].

\vspace{-0.7cm}

\end{document}